\documentclass[12pt,thmsa]{article}
\usepackage{aip}
\input{tcilatex}
\begin{document}

\author{E. Elizalde and M. Tierz \\
%EndAName
Instituto de Ciencias del Espacio (ICE/CSIC), \\
Institut d'Estudis Espacials de Catalunya (IEEC/CSIC), \\
Edifici Nexus, Gran Capit\`{a}, 2-4, 08034 Barcelona, Spain.}
\date{}
\title{Multiplicative anomaly and zeta factorization}
\maketitle

\begin{abstract}
Some aspects of the multiplicative anomaly of zeta determinants are
investigated. A rather simple approach is adopted and, in particular, the
question of zeta function factorization, together with its possible relation
with the multiplicative anomaly issue is discussed. We look primordially
into the zeta functions instead of the determinants themselves, as was done
in previous work. That provides a supplementary view, regarding the
appearance of the multiplicative anomaly. Finally, we briefly discuss
determinants of zeta functions that are not in the pseudodifferential
operator framework.
\end{abstract}

\section{Introduction}

A pseudodifferential operator ($\Psi$DO) $A$ of order $m$ on a manifold $%
M_{n}$ is defined through its symbol $a(x,\xi )$, which is a function
belonging to the space $S^{m}(\mbox{\bf R}^{n}\times \mbox{\bf R}^{n})$ of $%
\mbox{\bf C}^{\infty }$ functions such that for any $\alpha ,\beta $ there
exists a constant $C_{\alpha ,\beta }$ so that $\left\vert \partial _{\xi
}^{\alpha }\partial _{x}^{\beta }a(x,\xi )\right\vert \leq C_{\alpha ,\beta
}(1+|\xi |)^{m-|\alpha |}$. The definition of $A$ is given (in the
distribution sense) by 
\begin{equation}
Af(x)=(2\pi )^{-n}\int e^{i<x,\xi >}a(x,\xi )\hat{f}(\xi )\,d\xi ,
\end{equation}
where $f$ is a smooth function ($f\in \mathcal{S}$) and $\hat{f}$ \ its
Fourier transform. When $a(x,\xi )$ is a polynomial in $\xi $ one gets a
differential operator but, in general, the order $m$ can be even complex.
For $A$ a positive-definite elliptic $\Psi $DO of positive order $m\in %
\mbox{\bf R}$, acting on the space of smooth sections of an $n$-dimensional
vector bundle $E$ over a closed, $n$-dimensional manifold $M$, the zeta
function is defined as 
\begin{equation}
\zeta _{A}(s)=\mbox{tr}\ A^{-s}=\sum_{j}\lambda _{j}^{-s},\qquad \mbox{Re}\
s>\frac{n}{m}\equiv s_{0}.
\end{equation}
Here $s_{0}$ is called the abscissa of convergence of $\zeta _{A}(s)$, which
is proven to have a meromorphic continuation to the whole complex plane $%
\mbox{\bf C}$ (regular at $s_{0}$), provided that $A$ admits a spectral cut: 
$L_{\theta }=\left\{ \lambda \in \mbox{\bf C};\mbox{Arg}\,\lambda =\theta
,\theta _{1}<\theta <\theta _{2}\right\} $, $\mbox{Spec}\,A\cap L_{\theta
}=\emptyset $ (the Agmon-Nirenberg condition).

The Wodzicki (or noncommutative) residue \cite{wod1} is the only extension
of the Dixmier trace to $\Psi $DOs which are not in $\mathcal{L}^{(1,\infty
)}$. Even more, it is the only trace one can define in the algebra of $\Psi $%
DOs up to a multiplicative constant, and is given by the integral 
\begin{equation}
\mbox{res}\ A=\int_{S^{*}M}\mbox{tr}\ a_{n}(x,\xi )\,d\xi ,  \label{wr1}
\end{equation}
with $S^{*}M\subset T^{*}M$ the co-sphere bundle on $M$ (some authors a
coefficient in front of the integral). If dim $M=n=-$ ord $A$ ($M$ compact
Riemann, $A$ elliptic, $n\in \mbox{\bf N}$) it coincides with the Dixmier
trace, and one has \cite{wod1} 
\begin{equation}
\mbox{Res}_{s=1}\zeta _{A}(s)=\frac{1}{n}\,\mbox{res}\ A^{-1}.
\end{equation}
However, the Wodzicki residue continues to make sense for $\Psi $DOs of
arbitrary order and, even if the symbols $a_{j}(x,\xi )$, $j<m$, are not
invariant under coordinate choice, the integral in (\ref{wr1}) is, and
defines a trace. In particular, the residua of the poles of the extended
definition of zeta function %(not involving analytical continuation)    
to operators of complex order are also given by the noncommutative residue. 
% Res$_{s=s_k} \zeta_A^{KV} (s) =    
% \frac{1}{\mbox{ord}\, A}$  res $A^{-s_k}$.    

It is well known that the study of zeta functions is central ---at least at
a basic level, the one needed in fact in usual applications to physics \cite
{Vassilevich:2003xt,eli2}--- for the issue of giving a sense to the
definition of determinant of a $\Psi $DO (see Ref. $4$ for the actual state
of the art of this concept). This definition goes back to Ray and Singer 
\cite{rs1}: for an operator $A$ with spectrum $\lambda _{i},i\in I$ (here $I$
needs not be discrete, it can be a multi-index made up of parts of different
nature), formally 
\begin{equation}
\det A=\prod_{i\in I}\lambda _{i}=\exp \left( \sum_{i\in I}\log \lambda
_{i}\right) .
\end{equation}
But from the definition of the zeta function 
\begin{equation}
\zeta _{A}(s)=\sum_{i\in I}\lambda _{i}^{-s},
\end{equation}
it turns out that 
\begin{equation}
\zeta _{A}^{\prime }(0)=-\sum_{i\in I}\log \lambda _{i}.
\end{equation}
It is most natural then to define (as Ray and Singer did) the determinant of 
$A$ by means of the zeta function as \cite{rs1} 
\begin{equation}
{\det }_{\zeta }A\equiv \exp \left[ -{\zeta _{A}}^{\prime }(0)\right] ,
\end{equation}
Note that this is a \textit{definition}, since the above manipulations are
formal as long as the convergence properties of the expressions at hand are
not fully specified, in accordance with the theorem at the beginning of this
section. This is taken care of by the analytical continuation provided in
the definition of the zeta function of $A$.

The definition of the determinant $\det_{\zeta }A$ only depends on the
homotopy class of the spectral cut for $A$ (see above). And one has the
following (very useful) asymptotic expansion for the heat kernel, 
\begin{equation}
\hspace{-7mm}\mbox{tr}_{\ t\downarrow 0}\ e^{-tA}=\hspace{-5mm}\sum_{\lambda
\in \mbox{Spec}\,A}^{\prime }\hspace{-2mm}e^{-t\lambda }\sim \alpha
_{n}(A)+\sum_{n\neq j\geq 0}\alpha _{j}(A)t^{-s_{j}}+\sum_{k\geq 1}\beta
_{k}(A)t^{k}\ln t,  \label{asy1}
\end{equation}
where 
\begin{eqnarray}
&&\hspace{-6mm}\alpha _{n}(A)=\zeta _{A}(0),\qquad \alpha _{j}(A)=\Gamma
(s_{j})\mbox{Res}_{s=s_{j}}\,\zeta _{A}(s),\ s_{j}\notin -\mbox{\bf N},
\label{asy2} \\
&&\hspace{-16mm}\alpha _{j}(A)=\frac{(-1)^{k}}{k!}\left[ PP\zeta
_{A}(-k)+\psi (k+1)\mbox{Res}_{s=-k}\,\zeta _{A}(s)\right] ,\ s_{j}=-k,\
k\in \mbox{\bf N},  \nonumber \\
&&\beta _{k}(A)=\frac{(-1)^{k+1}}{k!}\mbox{Res}_{s=-k}\,\zeta _{A}(s),\ \
k\in \mbox{\bf N}\backslash \{0\}.  \nonumber
\end{eqnarray}

This paper is organized as follows: in the next section we give a short but
rather self-contained introduction to the appearance of the multiplicative
anomaly \cite{Mult,elicmp,ecvz} of zeta determinants. In Sec. $\mathrm{III},$
we point out certain particularities of this anomaly by presenting two very
different cases: On the one hand, a new and rather general condition that
guarantees the absence of anomaly and, in sharp contrast, a quite particular
and very simple case where the anomaly is already non-zero. Then, in Sec. $%
\mathrm{IV},$ extending and complementing previous work on this subject \cite
{elicmp,ecvz}, we pay attention to the product of zeta functions rather than
its associated determinants. This lead to the consideration of $\det
(B\otimes C)$ instead of $\det (BC),$ with $B$ and $C$ two arbitrary
operators. Thus, from that point of view, we are able to obtain rather
simple new expressions for determinants, mainly thanks to the strong
property of factorization of the zeta function. In the last section we
present, in a somewhat more qualitative way, the relationship between all
the previous concepts, multiplicative anomaly and zeta factorization, with
the appearance of complex poles in the zeta function and other zeta
functions that do not belong to the pseudodifferential operator framework.
In the Appendix, due to the implementation in a regularization context, we
investigate further the topic of zeta function factorization, presenting
results, mainly from Number Theory, with two opposite points of view: the
construction of Dirichlet $L$ functions from multiplication of simple zeta
functions on one hand, and the decomposition of a zeta function in terms of
simpler factors on the other hand. This leads to some physical
interpretation for the associated heat kernel that we briefly discuss.

\section{Appearance of the multiplicative anomaly}

It may seem clear that, if we have a product of two commuting operators, 
\begin{eqnarray}
{\det }_{\zeta }(AB) &=&\exp \left[ \sum_{i\in I}\log (\lambda _{i}\mu
_{i})\right] =\exp \left[ \sum_{i\in I}(\log \lambda _{i}+\log \mu
_{i})\right]  \nonumber \\
&=&\exp \left[ \sum_{i\in I}\log \lambda _{i}+\sum_{i\in I}\log \mu
_{i}\right] =\exp \left[ \sum_{i\in I}\log \lambda _{i}\right] \,\exp \left[
\sum_{i\in I}\log \mu _{i}\right]  \nonumber \\
&=&{\det }_{\zeta }A\ {\det }_{\zeta }B.
\end{eqnarray}
But this is \textit{not} true, and \textit{only one} of these steps fails to
be true. Below we provide some specific examples to help the reader
understand where the problem is.

Actually, very much related with this is the fact that the zeta function
trace 
\begin{equation}
\mbox{tr}_{\zeta }A=\sum_{i\in I}\lambda _{i}=\zeta _{A}(s=-1)
\end{equation}
fails to satisfy the additive property: in general 
\begin{equation}
\mbox{tr}_{\zeta }(A+B)\neq \mbox{tr}_{\zeta }A+\mbox{tr}_{\zeta }B,
\end{equation}
for, again, this is a regularized trace (involves analytical continuation)
which is used with non trace-class operators (see also Ref. $4$ for the
general definition of the trace).

As an example, consider the following commuting linear operators in an
infinite-dimensional space, given in diagonal form by: 
\begin{equation}
A_{1}=\mbox{diag\ }(1,2,3,4,\ldots ),\qquad A_{2}=\mbox{diag\ }%
(1,1,1,1,\ldots ),
\end{equation}
and their sum 
\begin{equation}
A_{1}+A_{2}=\mbox{diag\ }(2,3,4,5,\ldots ).
\end{equation}
The corresponding $\zeta $-traces are easily obtained: 
\begin{eqnarray}
&&\mbox{tr}_{\zeta }A_{1}=\zeta _{R}(-1)=-\frac{1}{12},\qquad \mbox{tr}%
_{\zeta }A_{2}=\zeta _{R}(0)=-\frac{1}{2},  \nonumber \\
&&\mbox{tr}_{\zeta }(A_{1}+A_{2})=\zeta _{R}(-1)-1=-\frac{13}{12},
\end{eqnarray}
$\zeta _{R}$ being the Riemann zeta function. The last trace has been
calculated according to the rules of infinite series summation (see, e.g.,
Hardy \cite{hardy}). We observe that 
\begin{equation}
\mbox{tr}_{\zeta }(A_{1}+A_{2})-\mbox{tr}_{\zeta }A_{1}-\mbox{tr}_{\zeta
}A_{2}=-\frac{1}{2}\neq 0.
\end{equation}

Unlike for ordinary, finite dimensional determinants, for which we have the
property: $\det (AB)=\det (A)\det (B),$ for zeta determinants one rather has
to consider, in general, an additional piece (called \textit{anomaly} or 
\textit{defect}). It is usually written as 
\begin{equation}
a(A,B)=\ln \frac{\det (AB)}{\det (A)\det (B)}
\end{equation}
or 
\begin{equation}
a\left( A,B\right) =\zeta _{A}^{\prime }\left( 0\right) +\zeta _{B}^{\prime
}\left( 0\right) -\zeta _{AB}^{\prime }\left( 0\right) .
\end{equation}
Thus the anomaly $a(A,B)$ will vanish if the derivatives at $s=0$ of the
respective zeta function satisfy the additive property. There is an explicit
expression, due to Wodzicki, for $a\left( A,B\right) ,$ that simplifies
enormously the calculation of the multiplicative anomaly in many cases \cite
{wod1}.

\section{Understanding zeta traces and zeta determinants}

There exist many examples of simple cases with and without multiplicative
anomaly \cite{elicmp,ecvz}. We give now a condition that guarantees its
absence. Consider the two following zeta functions: 
\begin{equation}
\zeta _{A}\left( s\right) =\sum_{i}\lambda _{i}^{-s},  \label{noanom1}
\end{equation}
\begin{equation}
\zeta _{B}\left( s\right) =\sum_{i}\left( c\lambda _{i}^{\alpha }\right)
^{-s}=c^{-s}\zeta _{A}\left( \alpha s\right) ,\ \mbox{with}\ c,\alpha \in %
\mbox{\bf R}.  \label{noanom2}
\end{equation}
The zeta function associated with the product of the eigenvalues is 
\begin{equation}
\zeta _{AB}\left( s\right) =\sum_{i}\left( c\lambda _{i}^{\alpha +1}\right)
^{-s}=c^{-s}\zeta _{A}\left( \left( \alpha +1\right) s\right) ,  \label{dem1}
\end{equation}
and thus 
\begin{equation}
\zeta_{AB}^{\prime}(s) = c^{-s} \left[ -\ln c \zeta_A ((\alpha +1)s) +
(\alpha +1) \zeta_A^{\prime}((\alpha +1)s) \right].
\end{equation}
Performing the substitution $s=0,$ we have that, 
\begin{equation}
\zeta_{AB}^{\prime}(0)= -\ln c \zeta_A (0) + (\alpha +1) \zeta_A^{\prime}(0)
=\zeta_A^{\prime}(0) + \zeta_B^{\prime}(0).  \label{dem2}
\end{equation}
Therefore, in spite of the fact that the two zeta functions are different,
their respective derivatives at zero are equal. This is enough to guarantee
the absence of the multiplicative anomaly, namely $a\left( A,B\right) =0$.
This is quite a general situation, since we have not fixed the $\lambda _{i}$
at all. We have only played with the relative difference between the spectra.

A rather different thing is to consider two spectra which are related by an 
\textit{additive} constant: 
\begin{equation}
\mu _{i}=\lambda _{i}+c.
\end{equation}
For simplicity, let us restrict our analysis to the specific example 
\begin{equation}
\lambda _{n}=n,\qquad \mu _{n}=n+1,\qquad n=1,2,3,...
\end{equation}
Thus 
\begin{equation}
\zeta _{A}(s)=\zeta _{R}(s),\qquad \zeta _{B}(s)=\zeta _{R}(s)-1,
\end{equation}
while the zeta function of the product is of Epstein type \cite{elicmp}: 
\begin{equation}
\zeta _{A}(s)=\sum_{n=1}^{\infty }(n^{2}+n)^{-s}=\sum_{n=1}^{\infty }\frac{%
\Gamma (n+s)\,2^{-2n}}{n!\,\Gamma (s)}\,\zeta _{H}(2(n+s),3/2).
\end{equation}
Thus 
\begin{equation}
\zeta _{A}^{\prime }\left( 0\right) +\zeta _{B}^{\prime }\left( 0\right)
=2\zeta _{R}^{\prime }\left( 0\right) =-\ln (2\pi ),
\end{equation}
while 
\begin{equation}
\zeta _{AB}^{\prime }\left( 0\right) =\sum_{n=1}^{\infty }\frac{2^{-2n}}{n}%
\,\zeta _{H}(2n,3/2),
\end{equation}
which are not equal. Numerically 
\begin{equation}
\zeta _{AB}^{\prime }\left( 0\right) =0.4417,\qquad \zeta _{A}^{\prime
}\left( 0\right) +\zeta _{B}^{\prime }\left( 0\right) =-1.8379,
\end{equation}
even the signs are different and the anomaly, in such a simple case, is
larger in absolute value than the individual results themselves, 
\begin{equation}
a=\zeta _{A}^{\prime }\left( 0\right) +\zeta _{B}^{\prime }\left( 0\right)
-\zeta _{AB}^{\prime }\left( 0\right) =-2.2796.
\end{equation}

Up to now, we have addressed and tried to explain the problem by looking
carefully in the various zeta functions involved in the process.
Nevertheless, we can gain a new insight into the multiplicative anomaly
issue through consideration of the factorizability properties of the
corresponding zeta functions, an analysis important by itself in, e.g.,
number theory.

\section{Zeta function factorizations and the multiplicative anomaly}

As explained, the main practical consequence about the existence of the
multiplicative anomaly is that, if, e.g. we want to compute 
\begin{equation}
\det A=\det (BC),
\end{equation}
from the (in principle simpler) determinants $\det B$ and $\det C,$ we have
to take also into account $a\left( A,B\right) .$ This is specially important
when different factorizations of $A$, say $A=BC$ and $A=B^{\prime }C^{\prime
}$, are alternatively considered \cite{ecvz}. We begin by introducing the
associated zeta functions that we would use in the computation of the factor
determinants, 
\begin{equation}
\zeta _{B}\left( s\right) =\sum_{i}\lambda _{i}^{-s},
\end{equation}
\begin{equation}
\zeta _{C}\left( s\right) =\sum_{j}\mu _{j}^{-s}.
\end{equation}
But, instead of applying the usual and direct procedure as before, here we
shall deal with the product of these two zeta functions, 
\begin{equation}
\zeta _{D}\left( s\right) =\zeta _{B}\left( s\right) \zeta _{C}\left(
s\right) =\sum_{i}\lambda _{i}^{-s}\sum_{j}\mu _{j}^{-s}.
\end{equation}
Note that this is the zeta function of an operator, $D$, which is different
from the previous $A.$ Actually, $D=B\otimes C$, as is immediate to realize.
In fact, from 
\begin{equation}
\zeta _{B\otimes C}(s)=\sum_{i,j}\left( \lambda _{i}\mu _{j}\right)
^{-s}=\sum_{i}\lambda _{i}^{-s}\sum_{j}\mu _{j}^{-s},\quad \mbox{Re }s>%
\mbox{max }\{\alpha ,\beta \},
\end{equation}
being $\alpha ,\beta ,$ the abscissas of convergence of the individual
series, and owing to the uniqueness of the asymptotic continuation to the
rest of the complex plane, it turns out that 
\begin{equation}
\zeta _{B\otimes C}(s)=\zeta _{B}(s)\zeta _{C}(s).
\end{equation}
In particular, 
\begin{equation}
\zeta _{B\otimes C}(s=-1)=\zeta _{B}(s=-1)\zeta _{C}(s=-1),
\end{equation}
that is 
\begin{equation}
\mbox{tr}_{\zeta }(B\otimes C)=\mbox{tr}_{\zeta }B\ \mbox{tr}_{\zeta }C,
\end{equation}
which extends the corresponding property known to hold in finite dimensions.

Now, consider the respective determinants. Recall, to begin with, that in
the finite case we have 
\begin{equation}
\det (B\otimes C)=(\det B)^{\mbox{dim }C}\,(\det C)^{\mbox{dim }B},
\end{equation}
where the dimensions refer to the spaces where the respective operators act.
We will now prove that this equation is maintained in the infinite
dimensional situation (we will drop the $\zeta $ label from the
determinants, from now on). In fact, we have (recall that $\zeta _{B}(0)$ is
the zeta \textit{regularized dimension} of the space in which $B$ acts, and
same for the rest): 
\begin{equation}
\det B=\exp [-\zeta _{B}^{^{\prime }}(0)],\qquad \det C=\exp [-\zeta
_{C}^{^{\prime }}(0)],
\end{equation}
\begin{equation}
\det D=\exp [-\zeta _{D}^{^{\prime }}(0)]=\exp [-\zeta _{B}^{\prime
}(0)\zeta _{C}(0)-\zeta _{B}(0)\zeta _{C}^{^{\prime }}(0)],
\end{equation}
and we thus see, that 
\begin{equation}
\det (D)=(\det B)^{\zeta _{C}(0)}(\det C)^{\zeta _{B}(0)}.
\end{equation}
In the particular case when $\zeta _{B}(s)$ and $\zeta _{C}(s)$ have the
same value at zero (the two operators act on a space of the same dimension), 
\cite{footnote} $\zeta _{B}(0)=\zeta _{C}(0)\equiv \widetilde{\zeta }(0)$,
we get 
\begin{equation}
\det (D)=(\det B\det C)^{\widetilde{\zeta }(0)}
\end{equation}
\qquad \ \qquad \ \ \ 

We have thus shown that the computation of $\det B\det C$ is, in a way, as
close to that of det $(B\otimes C)$ as it is to that of det $(BC)$, provided
when both operators act on the same space and can be multiplied. In fact,
the determinant of their tensor product is given in terms of the product of
the determinants of the individual operators by introducing the regularized
dimension of the space where they act. Formally, it is a kind of exponential
anomaly. But notice that this is actually no anomaly, since the exponent is
constant (e.g., it does not depend on the particular operators $B$ and $C$
chosen) and it is always equal to the regularized dimension of the space (as
it should). When $\zeta _{B}(0)=\zeta _{C}(0)=\widetilde{\zeta }(0),$\ let
us compare in more detail the two expressions: the one for the
multiplicative anomaly 
\begin{equation}
\det (BC)=\det B\det C\,e^{a(B,C)},
\end{equation}
with the other for the exponential anomaly, thus: 
\begin{equation}
\exp (a(B,C))=\frac{\det (BC)}{\det B\det C}=\frac{\det A}{(\det D)^{\frac{1%
}{\zeta (0)}}}.
\end{equation}
This equation seems somewhat artificial, no wonder since it links two non
directly related quantities, as explained above. It can nevertheless be
useful in practical determinations of the multiplicative anomaly.

\subsubsection{Some consequences and examples. }

In general, if one is dealing with factorizations of the type: 
\begin{equation}
\zeta _{A}(s)=\prod\limits_{i}\zeta _{A_{i}}(s),
\end{equation}
the determinants are related as $\det A=\prod\limits_{i}\left( \det
A_{i}\right) ^{\prod\limits_{j\neq i}\zeta _{j}(0)}.$ This can be useful for
the computation of determinants of multidimensional zeta functions, once its
factorization is known. For a general $m-$dimensional zeta function, we can
write its factorization as: $\zeta (s)=\prod\limits_{i}^{m}\zeta
_{i}^{d_{i}}(s)$ where $d_{i}$ specifies the dimension of the zeta function,
with $m=\sum\limits_{i}d_{i}$ .

A number of different examples can be worked out. For instance, if the zeta
functions factors are zero at the origin, then the associated
multidimensional determinant is one. This is what happens, for example, for
the product of harmonic oscillators, 
\begin{equation}
\prod\limits_{n_{1}=0}^{\infty }\cdot \cdot \cdot
\prod\limits_{n_{k}=0}^{\infty }\left( n_{1}+\frac{1}{2}\right) \cdot \cdot
\cdot \left( n_{k}+\frac{1}{2}\right) =1.  \label{418}
\end{equation}
Actually, with little more effort a more general case can be considered, 
\begin{equation}
\lambda _{n_{1}\ldots n_{k}}=(n_{1}+c_{1})\cdots (n_{k}+c_{k}),\qquad
n_{1},\ldots ,n_{k}=0,1,2,3,\ldots
\end{equation}
Here 
\begin{equation}
\zeta (s)=\prod_{j=1}^{k}\zeta _{j}(s),\qquad \zeta _{j}(s)\equiv \zeta
_{H}(s,c_{j}).
\end{equation}
Recalling that 
\begin{equation}
\zeta _{H}(0,c_{j})=\frac{1}{2}-c_{j},\qquad \zeta _{H}^{\prime
}(0,c_{j})=\ln \Gamma (c_{j})-\frac{1}{2}\ln (2\pi ),
\end{equation}
we get 
\begin{equation}
\det A=\prod_{j=1}^{k}\det (A_{j})^{\prod_{i\neq j}\zeta
_{H}(0,c_{j})}=\prod_{j=1}^{k}\left( \frac{\sqrt{2\pi }}{\Gamma (c_{j})}%
\right) ^{\prod_{i\neq j}(1/2-c_{i})},
\end{equation}
which reduces to the expression above, Eq. (\ref{418}), in the particular
example considered. This is a nice result of the regularization method.

A second example is the case of a multiple factorization, $\zeta ^{\left(
N\right) }(s)=\prod\limits_{i=1}^{N}\zeta _{i}(s)$, in which at least one of
the zeta functions evaluated at the origin is zero [without loosing
generality let us choose $\zeta _{1}(0)=0$]. Then, the determinant
associated with $\zeta ^{\left( N\right) }(s)$ is just 
\begin{equation}
\left( e^{-\zeta _{1}^{^{\prime }}(0)}\right) ^{\prod\limits_{i=2}^{\infty
}\zeta _{i}(0)},
\end{equation}
that is, the determinant of the zeta function which is zero at the origin,
exponentiated with the product of the other zeta functions at zero.
Different situations of this type could be discussed.

\section{Beyond $\Psi $DOs: the case of complex poles}

In this concluding section, we want to comment on the appearance, in some
important situations, of complex poles, and on its relationship with the
multiplicative anomaly and with factorizations. We begin by paying some
attention to the anomaly free case $\left( \ref{noanom1}\right) $ and $%
\left( \ref{noanom2}\right) $ studied in Sec. $\mathrm{III}.$ This case
corresponds to two commuting operators, for which there is a simple
expression for the multiplicative anomaly, due to Wodzicki \cite{wod1}, 
\begin{equation}
a\left( A,B\right) =\frac{\mbox{res }\left[ \left( \ln \left(
A^{b}B^{-a}\right) \right) ^{2}\right] }{2ab\left( a+b\right) },
\end{equation}
where $a>0$ and $b>0$ are the orders of $A$ and $B$ respectively.

In spite of the generality of $\left( \ref{noanom1}\right) $ and $\left( \ref
{noanom2}\right) $, it is clear that this is not the most general case
inside the class of commuting operators. To begin with, the fact that one is
a function of the other is a sufficient but not a necessary condition for
the commutation of the operators (think of the operators involved in the
quantum mechanics of the hydrogen atom, for example). In addition, one may
also argue that a more general function than $\mu _{n}=f\left( \lambda
_{n}\right) =c\lambda _{n}^{\alpha }$ may be considered as well. For
example, an exponential function $f\left( \lambda _{n}\right) =\exp \left(
\lambda _{n}\right) .$ It can be readily seen that with such a choice we are
outside the realm of pseudodifferential operators. For instance, just with $%
\lambda _{n}=n,$ then $\mu _{n}=\exp \left( n\right) ,$ and then the
associated zeta function is a geometric series: 
\begin{equation}
\zeta _{B}\left( s\right) =\sum_{n=1}^{\infty }\mathrm{e}^{-ns}=\frac{1}{%
\mathrm{e}^{s}-1},
\end{equation}
giving rise to infinitely many complex poles. Nevertheless, this spectra is
indeed physical, as shown in Ref. $11$, and related to $q$-deformations \cite
{Tierz:2003hk} and to fractal geometry \cite{Lap} as well. Thus, it is also
rather reasonable to expect that associated regularized expressions (such as
determinants) may be of physical interest as well. In principle, one can
proceed identically -depending on the precise meromorphic structure of the
corresponding zeta function- with the formal definition. Likewise, note that
the case $(\ref{noanom1})-(\ref{noanom2}),$ not only holds for $\alpha \in $ 
\textbf{R}, but also for $\alpha \in $ \textbf{C}, as can be readily seen
from $(\ref{dem1})-(\ref{dem2})$. Therefore, it is still anomaly free but
notice that a complex $\alpha $ introduces complex poles (just as a simple
example, consider $\mu _{n}=n,$ then the complex $\alpha $ rotates the pole
at $s=1$ to $s=\alpha ^{-1}$).

There are other circumstances where we are outside the pseudodifferential
operator framework, but there is still interest in the short time
asymptotics of the heat kernel or in zeta determinants. This is exactly the
case, for instance, when considering heat kernels in noncommutative spaces 
\cite{Vassilevich:2003yz} and when studying products of prime numbers \cite
{ric}, respectively. Indeed, the zeta function associated to the prime
numbers is known \cite{Land},

\begin{equation}
\mathcal{P}\left( s\right) =\sum_{p}p^{-s}=\sum_{n=1}^{\infty }\frac{\mu
\left( n\right) }{n}\log \zeta \left( ns\right) ,
\end{equation}
where $p$ are the prime numbers and $\mu \left( n\right) $ is the M\"{o}bius
function. Note that this function has a rich pattern of logarithmic
singularities in the complex plane but still the associated \textit{%
determinant} is of interest and actually follows directly from the
derivative of $\mathcal{P}\left( s\right) $ \cite{ric}. Additionally, in
this type of regularized products, one can look at multiplicative anomalies
as well. For example, following Ref. $14$, one can consider the Euler
product representation of Riemann's zeta function, 
\begin{equation}
\zeta \left( s\right) =\frac{\prod_{p}p^{s}}{\prod_{p}p^{s}-1},\quad %
\mbox{Re}\ s>1.
\end{equation}
This expression, considered together with $\mathcal{P}\left( s\right) ,$
gives rise to the following result \cite{ric}: 
\begin{equation}
\prod_{p}\left( p-1\right) =0\quad \text{and }\prod_{p}\left( p^{2}-1\right)
=48\pi ^{2},
\end{equation}
and the appearance of a multiplicative anomaly is manifest.

\bigskip

\noindent\textbf{Acknowledgments}

E.E. is indebted to the Mathematics Department, MIT, especially Dan
Freedman,
for the warm hospitality. One of the author's (E.E.) investigation has
been partially supported by DGI/SGPI (Spain), project BFM2000-0810, and by
CIRIT (Generalitat de Catalunya), contract 2001SGR-00427. M.T. is indebted
to Michel Lapidus for
many useful and interesting discussions on zeta functions, and also for warm
hospitality at University of California at Riverside.

%\newpage   

\appendix 

\section{Remarks on zeta factorizations.}

We have seen how the discussion of the multiplicative anomaly of
determinants, has lead us, in a natural way, to the construction of a zeta
function from the product of other zeta functions. Generically, and
following the previous notation, let us envisage 
\begin{equation}
\zeta _{D}\left( s\right) =\zeta _{B}\left( s\right) \zeta _{C}\left(
s\right) =\sum_{i}\lambda _{i}^{-s}\sum_{j}\mu _{j}^{-s}.
\end{equation}
This turns out to be an important construction in number theory. Actually,
even with the simplest zeta functions as factors, important and
sophisticated $\zeta _{D}\left( s\right) $ are obtained. For example, with
the Riemann zeta function itself. In fact, from the Euler product of the
Riemann zeta function, we know that it has local factors of degree $1$ at
each prime, while automorphic $L$ functions have local factors of degree $2$
at almost all places \cite{Iwa}. This suggests that we can denote such
product as 
\begin{equation}
L\left( s\right) =\zeta _{R}\left( s\right) \zeta _{R}\left( s-k+1\right) ,
\label{L}
\end{equation}
with $k\geq 2$. In Ref. $16$ it is shown that the $L$ function is actually 
\begin{equation}
L\left( s\right) =\sum_{n=1}^{\infty }\sigma _{k-1}\left( n\right) n^{-s},
\end{equation}
where $\sigma _{k}$ is the arithmetic function (the generalized divisor
function, or sum over all the divisor of $n$ to some power), given by 
\begin{equation}
\sigma _{k}\left( n\right) =\sum_{d\mid n}d^{k}.  \label{div}
\end{equation}
This appears naturally in the Chowla-Selberg formula and its generalizations 
\cite{elicmp}.

This shows, in close relationship with the preceding section, how the
product of even the simplest of the zeta functions lead to an interesting
object by the process considered above, often with important arithmetic
properties (and some of these $L$ functions are useful in analytical
approaches to the study of algorithms \cite{Flaj,Flaj2}). Even more, in
general, increasingly complex $L$ functions are very often constructed or
represented by a generic product of simpler $L$ functions \cite{Iwa}.

Nevertheless, it seems apparent that instead of exploiting the useful idea
of constructing zeta functions, it may also be worth to look at this
relation from the other side, that is, as a decomposition of the zeta
function on the l.h.s. into several factors. To illustrate the approach for
zeta functions, let us just take into account the two simple examples
considered in detail in Ref. $19$. The zeta function 
\begin{equation}
\zeta (s)={\sum_{m,n\in \mbox{\bf Z}^{2}}}^{\prime }(m^{2}+n^{2})^{-s},
\end{equation}
with the summation extended over all pairs $(m,n)\neq (0,0)$ in $%
\mbox{\bf   
Z}^{2},$ can be expressed as 
\begin{equation}
\zeta (s)=4\zeta _{R}(s)\cdot L(\chi _{4},s),
\end{equation}
where $\zeta _{R}(s)$ is the Riemann zeta function and $L(\chi _{4},s)$ is
the Dirichlet zeta function corresponding to the character $\chi _{4}.$
Another interesting factorization is the following one, for a different
particular case of the two-dimensional Epstein zeta function 
\begin{equation}
\zeta (s)={\sum_{m,n\in \mbox{\bf Z}^{2}}}^{\prime
}(m^{2}+mn+n^{2})^{-s}=6\zeta _{R}(s)\cdot L(\chi _{3},s).
\end{equation}
Once again, we see the natural appearance of $L$ functions, whose
determinants are of much interest as well (mainly in a number theoretical
context; see Ref. $20$ for a review).

These factorizations are particular cases of a more general situation coming
from algebraic considerations in number theory \cite{Iwa}. Very general
statements are not always possible, but let us compare the previous with the
classical results (due to Dirichlet) concerning primitive quadratic forms of
any determinant, 
\begin{equation}
Q\left( x\right) =ax_{1}^{2}+bx_{1}x_{2}+cx_{2}^{2},\qquad \left(
a,b,c\right) =1
\end{equation}
(the parentheses meaning here maximum common divisor), with $D=-\det
Q=b^{2}-4ac<0$ (the discriminant of $Q$), and 
\begin{equation}
\chi _{D}\left( d\right) =\left( \frac{D}{d}\right) .
\end{equation}
Then, for $n>0,$ $\left( n,D\right) =1$, the character sum 
\begin{equation}
r\left( n;D\right) =\omega _{D}\sum_{d\mid n}\chi _{D}\left( d\right)
\end{equation}
gives the number of all representations of $n$ by representatives of forms
of all classes of discriminant $D$. Here $\omega _{D}$ stands for the number
of automorphs: 
\begin{equation}
\omega _{D}=\left\{ 
\begin{array}{ll}
6, & \mbox{if}\ D=-3, \\ 
4, & \mbox{if}\ D=-4, \\ 
2, & \mbox{if}\ D<-4.
\end{array}
\right.
\end{equation}
Notice how the discriminant gives the right character for the $L$ function
and the number of automorphs the right prefactor in the previous example of
factorization. Nevertheless, we must point out that these previous examples
and the posterior discussion looks so simple, due to the fact that the
examples correspond to discriminants $D$ for which the class number $h\left(
D\right) $ (the number of equivalence classes of primitive binary quadratic
forms) is one.

\subsection{Factorization at the level of the heat kernels.}

Now, we pay attention to the meaning of the zeta factorization at the level
of the respective associated heat kernels. Since $A^{-s}$ and $\exp \left(
-tA\right) $ are related by the following expression: 
\begin{equation}
A^{-s}=\frac{1}{\Gamma \left( s\right) }\int_{0}^{\infty }t^{s-1}\exp \left(
-tA\right) dt,
\end{equation}
then the zeta function is, up to a gamma function, the Mellin transform of
the heat kernel. The interest of this expression, considered together with
the factorization property, is that it allows, in a probabilistic context,
the product of random variables to be directly performed in Mellin space (in
contrast to the better known case of the addition of variables, where
Fourier transform is used) \cite{Spri}. Therefore, the zeta factorization
implies also a product for the respective heat kernels, but a product in the
sense of probability theory, that is, the heat kernel (or in a number
theoretical context, the theta function) denotes the probability
distribution function of a random variable $X_{i},$ and then we have the
product $X=\prod_{i=1}^{n}X_{i}.$ Nevertheless, for this to be exactly
correct we should take into account the gamma function for each factor and
for the resulting zeta function. For example, in the case of a zeta function
with two factors: 
\begin{eqnarray}
\zeta \left( s\right) =\zeta _{1}\left( s\right) \zeta _{2}\left( s\right)
&\longrightarrow &\Gamma \left( s\right) \Gamma \left( s\right) \zeta \left(
s\right) =\Gamma \left( s\right) \zeta _{1}\left( s\right) \Gamma \left(
s\right) \zeta _{2}\left( s\right) \\
&\longrightarrow &K\left( t\right) \cdot \exp \left( -t\right) =K_{1}\left(
t\right) \cdot K_{2}\left( t\right) ,  \nonumber
\end{eqnarray}
where, in the last expressions, the products are in the sense explained
above, and we have used the fact that $\Gamma \left( s\right)
=\int_{0}^{\infty }t^{s-1}\exp \left( -t\right) dt.$ Thus, the necessary
introduction of gamma factors implies that we have to take into account
possible products of the main heat kernel with an exponential distribution.

This stochastic point of view seems both interesting from the mathematical
side, where a probabilistic interpretation of zeta and theta functions is of
interest \cite{bpy}, and also from a physics perspective, where products of
random variables very often constitute a role model of what is known with
the name of multiplicative or cascade processes \cite{sor}.

Last but not least, the factorization is potentially interesting from the
practical point of view in the asymptotic study of the trace of the heat
kernel (\ref{asy1}) and (\ref{asy2}). The contributions can be considered
separately, with the exception of the possible coincidence of poles or poles
and zeros. This fact introduces interesting phenomena that can be seen with
the following example. Consider the product of two Riemann zeta functions, 
\begin{equation}
\zeta \left( s\right) =\zeta _{R}\left( s\right) \zeta _{R}\left( s\right) ,
\end{equation}
which yield the well-known $L$ function, 
\begin{equation}
\zeta \left( s\right) =L\left( s\right) =\sum_{k=1}^{\infty }\frac{d\left(
k\right) }{k^{s}},
\end{equation}
with $d\left( k\right) $ the divisor function again. Note the consistency
with the previous case (\ref{L})-(\ref{div}$).$ The idea is now to construct
another zeta function from two very similar factors, 
\begin{equation}
\zeta _{\varepsilon }\left( s\right) =\zeta _{R}\left( s\left( 1+\varepsilon
\right) \right) \zeta _{R}\left( s\left( 1-\varepsilon \right) \right) ,
\end{equation}
with $\varepsilon >0$ a very small, real positive number. It seems that
these two zeta functions should be almost identical in the whole complex
plane, except for the fact that, in the first one, we have a double pole at $%
s=1,$ while the second has two simple poles at $s=\left( 1+\varepsilon
\right) ^{-1}$ and $s=\left( 1-\varepsilon \right) ^{-1}$, very close one to
the other for $\varepsilon $ small. Note that the point-like structure of a
pole allows to play that game. Now, from (\ref{asy1}) and (\ref{asy2}), it
is clear how different the $t\rightarrow 0$ expansion of the associated
trace of the heat kernel is, in the two cases. In the first case, we have 
\begin{equation}
\hspace{-7mm}\mbox{tr}\ e^{-tA_{1}}\sim -\frac{\log t}{t},\ t\downarrow 0,
\end{equation}
in sharp contrast with the second case, where 
\begin{equation}
\hspace{-7mm}\mbox{tr}\ e^{-tA_{2}}\sim \Gamma \left( \frac{1}{1+\varepsilon 
}\right) t^{-\left( \frac{1}{1+\varepsilon }\right) }+\Gamma \left( \frac{1}{%
1-\varepsilon }\right) t^{-\left( \frac{1}{1-\varepsilon }\right) },\
t\downarrow 0.
\end{equation}

We see that the case where the poles collide possesses a partition function
which is much larger the smaller the value of $t$ is (the classical limit).
Therefore, the associated partition functions differ considerably in the
classical limit. A deeper physical understanding of this phenomena seems to
be an interesting open question.


\begin{thebibliography}{99}
\bibitem{footnote}  Notice that this is an inescapable requirement if the
ordinary product of the two operators is to make any sense.
\end{thebibliography}
\end{document}